\documentclass[fleqn,usenatbib]{mnras}

\usepackage{newtxtext,newtxmath}

\usepackage{natbib}
\usepackage[T1]{fontenc}
\usepackage{cuted}
\DeclareRobustCommand{\VAN}[3]{#2}
\let\VANthebibliography\thebibliography
\def\thebibliography{\DeclareRobustCommand{\VAN}[3]{##3}\VANthebibliography}

\usepackage{graphicx}	
\usepackage{amsmath}	

\title[The Statistical Analysis Of The Galactic Open Clusters' Structure]{The Statistical Analysis Of The Galactic Open Clusters' Structure}

\author[Jin-Sheng Qiu et al.]{
Jin-Sheng Qiu,$^{1, 2}$\thanks{Corresponding author: Zhen Wan, Xu-Zhi Li ,Qing-Feng Zhu} 
\thanks{E-mail:qjs003@mail.ustc.edu.cn,zhen\_wan@ustc.edu.cn,lixuzhi@ustc.edu.cn, zhuqf@ustc.edu.cn}
Zhen Wan, $^{1, 2}$ 
Xu-Zhi Li,$^{3, 4}$
Qing-Feng Zhu,$^{1, 2}$
Lu-lu Fan,$^{1, 2, 5, 6}$
Xiao-Hui Xu,$^{1, 2}$
\protect\\ \normalfont \Large
Jun-Han Zhao,$^{1, 2}$
and Zhi-Yong Pu$^{1, 2}$ \\
$^{1}$Department of Astronomy, University of Science and Technology of China, Hefei 230026, China \\
$^{2}$School of Astronomy and Space Sciences, University of Science and Technology of China, Hefei 230026, China \\
$^{3}$School of Mathematics and Physics, Anqing Normal University, Anqing 246133, China \\
$^{4}$Institute of Astronomy and Astrophysics, Anqing Normal University, Anqing 246133, China \\
$^{5}$Deep Space Exploration Laboratory, Hefei 230088, People's Republic of China \\
$^{6}$College of Physics, Guizhou University, 550025 Guiyang, PR China
}

\pubyear{\the\year{}}

\begin{document}
\label{firstpage}
\pagerange{\pageref{firstpage}--\pageref{lastpage}}
\maketitle

\begin{abstract}
We present a systematic investigation of 1,481 Galactic open clusters (OCs) through the application of the Limepy  dynamical model, from which we derive the fundamental structural parameters  of OCs. We conduct the statistical analyses on the structural parameters with clusters’ ages and locations within the Milky Way. 
Our results reveal the higher  concentration in the cluster center is associated with the sharper truncation at the periphery of cluster, which is consistent with previous findings for globular clusters (GCs). We further find the systematic increase of the lower limit of clusters' half-mass radius ($R_h$) with age.  Our results also show that OCs located at larger vertical distances from the Galactic plane systematically display higher central concentrations. 
Our findings collectively suggest that the structural characteristics of OCs are shaped by both  intrinsic evolutionary processes and interactions with the Galactic environment. During the evolution of star clusters, the combined effects of mass segregation and tidal stripping  lead to the systematic pattern between  central concentration and outer truncation. Clusters of different ages and locations within the Milky Way undergo different evolutionary histories, resulting in correlations between the $R_h$ and age, as well as between central concentration and galactic location.
\end{abstract}




\section{Introduction}

The stars within an OC are generally considered to be formed almost at the same time, sharing the  similar initial environmental conditions \citep{2003Lada}. The properties of OCs exhibit differences in various aspects. For instance, the numbers of member stars, the ages, and the characteristic radii vary widely among different OCs from previous surveys \citep[e.g.][]{2002Dias,2013Kharchenko,2020Cantat-Gaudin,2019LiuPang,2022Hao, 2022He, 2023He, 2022Li, 2023Li, 2023Chi, 2024Cavallo}. In addition, the structural characteristics, such as central density and peripheral morphology, also vary among different OCs \citep[e.g.][]{2005Bonatto,Bisht_2020,2020Zhang,Pang_2021, 2022Pang,2022Tarricq}. To understand these variations, it is essential to systematically study a large sample of OCs, which enables the exploration of the potential systematic characteristics and connections in their structures and properties
\citep{2001Bergond,2003Heggie,2024Cantat-Gaudin}.

The structures of OCs can be influenced by internal evolution processes. For example, two-body relaxation plays an important role in reshaping cluster morphology on the relaxation timescale \citep{1958Spitzer,2001Bergond,2005Bonatto}. 
It causes mass segregation phenomenon, which shows that massive stars tend to sink towards the center of the cluster while low-mass stars are pushed towards the outer region of the cluster \citep{1985IAUS..113..427M,1987Spitzer, 2000Takahashi, 2005Bonatto,2019Krumholz}. For the cluster with 3 to 40  Myr, supernovae (SNe) in the OC produce high-speed gas that   escapes the  cluster. The SNe remnant such as neutron star or black hole can receive the kick velocity of several hundred $km\ s^{-1}$, which may escape the cluster as well \citep{2006Faucher,2019Krumholz}. These can cause the mass loss of cluster,  alter the internal gravitational potential and trigger cluster expansion
\citep{2019Krumholz}. 
 
The OCs   also experience the interactions with the Milky Way \citep{2001Bergond, 2005Bonatto, 2024Cantat-Gaudin}. For example, the tidal stripping  driven by the tidal forces can cause the outer stars of a  cluster to be stripped away \citep{Lamers_2005,2006Gieles,2019Krumholz}. The
continuous stripping may  produce the steeply truncated radial density profile in the cluster's outer regions, as demonstrated through the N-body simulations \citep{2016Zocchi}. In addition, observations show that the inner disk of the Milky Way (Galactocentric distances $R_{gc} < 6.5\ kpc$) is relatively deficient in older OCs \citep{2024Cantat-Gaudin}. It suggests that  the dense environment of the inner disk can lead to a high disruption rate for OCs \citep{2020Cantat-Gaudin,2024Cantat-Gaudin}.
Observations also show the scarcity of OCs beyond a few hundred parsecs from the Galactic plane. One of the explanations is that clusters far from the Galactic plane  experience strong vertical tidal shocks and are more susceptible to disruption when they cross the Galactic disk  \citep{2017Martinez,2019Krumholz, 2019Webb}. In addition, the encounters between OCs and giant molecular clouds (GMCs) in the Milky Way can induce impulsive perturbations on the clusters and cause the clusters to gain energy, further leading to the escape of member stars, the cluster expansion and even the dissolution of the cluster
 \citep {2006Gieles, 2008Binney, 2010Gieles,2019Krumholz, 2019Webb}. 

The quantification of star cluster structures plays a key role in investigating the influence of evolutionary processes. One widely used method is modeling their surface density profiles with a suitable distribution function (DF).
Numerous models have been employed, such as Woolley model \citep{1954Woolley}, King model \citep{1966King} and Wilson model \citep{1975Wilson}. \cite{2014Gomez-Leyton} and \cite{2015Gieles}  proposed that these models are special cases of the Limepy\footnote{https://github.com/mgieles/Limepy} (Lowered Isothermal) model. Limepy model is a class of star cluster structural models based on the lowered isothermal DF. Its core idea is to truncate the classical isothermal sphere model at the outskirt of the cluster, mimicking the effect of the external tidal field on the spatial extent of the stellar system \citep{2024Alvarez-Baena}. Compared to earlier models, the Limepy model provides a more detailed description of the stellar distribution near the escape energy of the cluster. Furthermore, Limepy is particularly effective in describing the phase-space density distribution of stars in tidally truncated and mass-segregated clusters across their entire evolutionary lifespan, thereby enabling a detailed investigation into the structural evolution of star clusters \citep{2024Alvarez-Baena}.
The Limepy model has been proven effective in fitting the observations of star clusters. For example, \cite{2019de-Boer} applied Limepy model to fit 81 globular clusters' surface density profiles and got their structural parameters. \cite{2024Alvarez-Baena} utilized Limepy model to analyze the structural properties of six old OCs in great detail.  



This paper applies the Limepy model for a large sample of OCs, thereby investigating the statistical relationships of their structures and other physical parameters. By using the Limepy model, we can achieve   comprehensive understanding of the OCs' structural properties. Our investigation specifically focuses on determining and quantifying the relationships between the structural parameters and the basic physical information of OCs, such as the ages and locations within the Milky Way. 

This work is organized as follows. In Section \ref{sec:2}, we introduce the  data of the clusters used for this research. In Section \ref{sec:3}, we describe our modified model, which combines the Limepy model and the uniform model for cluster members and field stars. In Section \ref{sec:4}, we explore the relationships between the structural parameters and derive the statistical regularities. Our conclusions and discussions are presented in Section \ref{sec:5}.

\section{Data} \label{sec:2}

With the release of Gaia Data Release 3 (DR3), we now have access to the  large and accurate astrometry dataset, including stellar R.A., decl., parallax, proper motions, and photometric magnitudes in the G, BP, and RP bands for more than a billion objects \citep{GaiaCollaboration2023}.
The rich dataset from Gaia DR3 ensures that the cluster members are as complete as possible, which greatly benefits the determination of cluster properties and the analysis of their structural characteristics \citep[e.g.,][]{2024Qiu, 2022Pang, 2023Long}.

A variety of methods are used to identify cluster member stars, but differentiating them from field stars in the outer region of the cluster is a  challenge.
To mitigate the biases introduced by member star selection, we adopt the model which incorporates both cluster members and foreground/background stars (see details in Section \ref{sec:3}). This method ensures a comprehensive description of the stellar  distribution within the field of the cluster,  reducing potential systematic errors associated with member star selection.
Correspondingly, we allow the existence of field stars within our dataset, instead of using the catalog of the selected cluster members (e.g., \cite{Cantat-Gaudin2020}, hereafter denoted as CG20). 

We select 1,481 OCs  from the CG20 catalog as our sample. We obtain the basic information of OCs from CG20, such as center positions, parallaxes, mean proper motions, radii $r_{50}$ (radius containing half the members) of clusters. The dataset for our study is collected within a sufficiently large range around the cluster centers to include a complete sample of member stars.
We initially adopt a search radius of 5 $r_{50}$ as the range for each cluster. The spatial distribution of the collected data is compared with the CG20 member list to verify that all cataloged members are encompassed within this radius.  
For clusters whose members extend beyond 5 $r_{50}$, the search area is extended gradually to include all cataloged members. The final adopted radius varies depending on the spatial extent of the CG20  members, typically ranging from
5 $r_{50}$ to 10 $r_{50}$,  but may differ for individual clusters.
A loose selection of the proper motion is further applied on the data to exclude stars that are confidently non-members. We allow the stellar sample in a cluster to have a velocity deviation of  $5\ \mathrm{km/s}$ around the mean cluster's velocity, as \cite{Cantat-Gaudin2020}. This range is larger than the classical OCs' intrinsic velocity deviation $1\ \mathrm{km/s}$ given by \cite{1989Girard, Mathieu2015}. This velocity deviation can be translated into a range of proper motion through the cluster parallax $\omega$, as the Eq ~\eqref{eq1}
\citep{Cantat-Gaudin2020, 2022Hao}.  Because of the larger proper motion uncertainty for the distant clusters, we set a lower limit of $1.5\ \mathrm{mas/yr}$ for proper motion range and the selection is larger than  previous works \citep{Cantat-Gaudin2020, 2022Hao}. We choose the stars within the proper motion range based on the mean proper motion from CG20. We also compare the selected data with the cluster members catalog CG20 in proper motion space to ensure the members in CG20 are fully covered by our dataset.

To further reduce the contamination from non-cluster members, we use a photometric criterion based on theoretical isochrones for clusters with available ages, extinctions and distance moduli from CG20. Using the CMD 3.8{\footnote{http://stev.oapd.inaf.it/cgi-bin/cmd\_3.8}} \citep{2012Bressan, 2013Marigo}, we obtain an isochrone ($G_0\ vs.\ (BP-RP)_0$) for each cluster assuming solar metallicity. We then select stars that lie within a 2.0 mag ($\pm1 $ mag) strip around the isochrone in the color-magnitude diagram as Eq \eqref{eq1_1}. For the subset of clusters ($\sim 110$) lacking ages and extinctions  in CG20, only the proper motion criterion is applied.
Additionally, we impose a limit of 22 mag on the apparent magnitudes G, BP, and RP band.

We note that Gaia DR3 may suffer from incompleteness in high-density regions, particularly in the center of GCs \citep{2019de-Boer}. However, for OCs analyzed in this work, we find no evidence of central incompleteness. Surface density profiles and spatial distributions show centrally concentrated structures without central depletion.

\begin{equation}
    \Delta\mu\ (mas/yr) =
    \begin{cases}
    \omega\ (mas) \times \frac{5\ (km/s)}{4.74}, \ \ \omega > 1.422\ mas\\
    1.5, \ \ \omega \leq 1.422\ mas
    \end{cases}
    \label{eq1}
\end{equation}

\begin{equation}
|(BP-RP)_{obs} - (BP-RP)_{0}| < 1.0\ mag,\ as\ G_{obs}=G_{0}
    \label{eq1_1}
\end{equation}

\section{Fitting The Limepy Model} \label{sec:3}

The Limepy model describes the  phase-space distribution of spherical stellar systems with different central concentrations and truncations. The isotropic Limepy model is characterized by three independent parameters, which are: (1) $W_0$: (also denoted as $\hat{\phi}(0)$) the boundary condition required for solving the dimensionless Poisson’s equation about gravitational potential function $\phi(r)$. A larger $W_0$ value indicates a more concentrated core of the cluster. (2) g: the parameter which regulates the sharpness of the truncation at the outskirt of cluster. A lower g value indicates a sharper truncation of the surface density. (3) $R_h$: the half-mass radius of a star cluster.

Our dataset for each OC includes both the member stars and field stars. We correspondingly develop a model that combines the Limepy model distribution with a uniform background of field stars, following a similar approach  presented by \cite{2016Zocchi}. With the cluster's center position and the heliocentric distance from \cite{Cantat-Gaudin2020}, we calculate the projection distances $R_i\ (pc)$  to the cluster center of both cluster member stars and field stars, from which we derive the surface  density distribution, and construct the likelihood function as  Eqs~\eqref{eq:eq1}--\eqref{eq:eq3}.

\begin{figure*}
\centering
\begin{minipage}{0.9\textwidth}
\begin{subequations}
\label{eq:all}
\begin{flalign}
    &P(r \sim r+\Delta r) = f(r) \times 2\pi r\Delta r \equiv F(r) \Delta r, 
    \quad \int^{R_{\mathrm{max}}}_{0} F(r)\,\mathrm{d}r=1 & \label{eq:eq1} \\[1em]
    &F(r) \equiv (1-A)\frac{\Phi(r)}{\int^{R_{\mathrm{max}}}_{0}\Phi(r)\, \mathrm{d}r}
    + A\frac{r}{\int^{R_{\mathrm{max}}}_{0} r\, \mathrm{d}r}, 
    \quad  \Phi(r) \equiv r\times\phi_{\mathrm{Limepy}}(r) & \label{eq:eq2} \\[1em]
    &\log\text{ Likelihood Function} \equiv \sum_{i} \log F(r)|_{r=R_{i}} \equiv \sum_{i} \log\left[(1-A)\frac{\Phi(r)}{\int^{R_{\mathrm{max}}}_{0}\Phi(r)\, \mathrm{d}r}
    + A\frac{r}{\int^{R_{\mathrm{max}}}_{0} r\, \mathrm{d}r}\right]\Bigg|_{r=R_{i}} & \label{eq:eq3}
\end{flalign}
\end{subequations}
\end{minipage}
\label{fig:formulas}
\end{figure*}

\begin{figure*}
   \centering
   \includegraphics[width=1\linewidth]{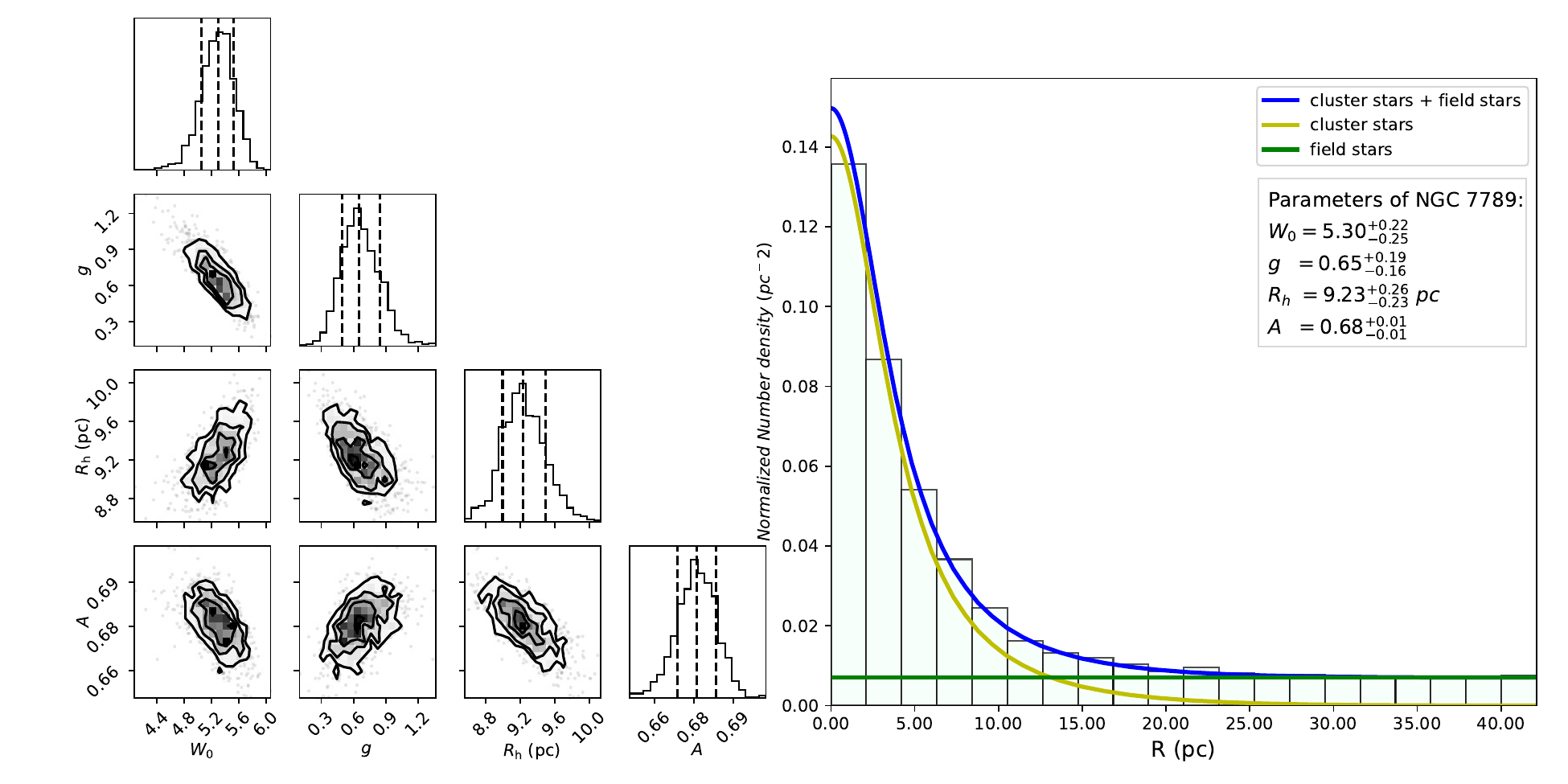}
    \caption{We take NGC 7789 as an example for our process. (Left-panel) The corner diagram of the MCMC result for four parameters in our model. (Right-panel) The compared results between the model obtained from MCMC and the original data. The histogram is original data which translates into the normalized number density distribution. The blue line is the model with the best-parameters from MCMC. The yellow line and green line represent cluster members and field stars, respectively.}
    \label{fig:1}
\end{figure*}

In the formulas, $P(r \sim r+\Delta r)$ is the probability of finding a
star in the projection distance $r \sim r+\Delta r$. $\phi_{Limepy}(r|W_{0}, g, R_{h})$ represents the Limepy distribution function, where $W_{0}$, $g$, $R_{h}\ (pc)$ are the parameters of Limepy model \citep{2015Gieles}. The parameter $A$ denotes the ratio of the number of field stars to the total number of stars within each cluster sample. $R_{max}\ (pc)$ is the maximum projection distance between the stars and the cluster center. The function $F(r) $ satisfies the following conditions from Eq~\eqref{eq:eq2}. First, the integral of $F(r) $ from 0 to $R_{max}$ is normalized. Second, the surface density distribution $f(r) $ ($\equiv \frac{F(r)}{2 \pi r}$) follows the form of the Limepy distribution function plus a constant (stands for uniform background of field stars). We can calculate the probability of finding a
star in the position $(r \sim r+\Delta r)$ by a given parameters of model. The likelihood function is the product of individual probabilities calculated for every star in the  dataset of the projection distances ($R_i$). We convert the likelihood function into the  logarithmic form for computational convenience as Eq~\eqref{eq:eq3}. The Monte Carlo Markov Chain (MCMC) method is applied to sample our model's  parameter space. The emcee, a PYTHON implementation of MCMC sampler, is used to sample the posterior probability distribution for the parameters \citep{emcee}.
Uniform prior distributions are adopted over the following ranges: $0 < W_0< 10$, $0 < g < 3.5$, $0 < R_h< 50 \ pc$, $0 < A< 1$. The MCMC sampling  runs for 8,000 steps to ensure that each parameter converges to a stable and reliable value. The left panel of Figure~\ref{fig:1} presents the corner diagram of the sampling result, taking NGC 7789  as an  example. 

Aiming to check the observational results with the model, we compare the surface density profile  $f(r)$ to the observation data. The right panel of Figure~\ref{fig:1} shows the result of NGC 7789 as an example. We divide the area of the  cluster into a series of
equal-width  concentric rings around the cluster center and  calculate the number density:  $\frac{N_i}{A_i}$ for the i-th concentric ring, where $A_i$ is the area of the i-th concentric ring. 
Both the surface density  profile $f(r) $ and the number density distribution in the right panel of Figure \ref{fig:1} are normalized. The blue line is the surface density function with the best-parameters model from MCMC, and the corresponding parameters are noted in the upper right corner. The yellow line and green line represent cluster members and field stars, respectively.

\section{Result} \label{sec:4}

We obtain the structural parameters for 1,481 OCs in our sample by using the model described by Eqs~\eqref{eq:eq1}--\eqref{eq:eq3}. For each cluster, we compare the the model surface density profile $f(r) $ of the corresponding parameters with the observed surface number density profile.

In our sample, we select the clusters for which our model fails to reproduce the observed profiles. The examples of those clusters are shown in Appendix \ref{appendixA}.  For about 160 OCs in our sample, we find that the field stars near these OCs are spatially inhomogeneous (e.g. Loden 1194 in Appendix \ref{appendixA}). This contradicts the assumption of uniformly distributed field stars in our model.
The structural parameter estimations for those OCs have no physical meaning and are excluded from the further statistical analysis. 
Among the remaining 1315 OCs, a part of OCs exhibit discrepancies between observations and models in the central regions.
We select those OCs with large deviations and apply a simple criterion: $|Observation - Model|\ /\ Observation > 0.25$ for the center bin, which is the first of 20 equal-width bins in clusters' radial distribution, to distinguish two groups of clusters: Group A  ($< 0.25$, 1160 OCs) and Group B ($> 0.25$, 155 OCs, e.g. ASCC 123 in Appendix \ref{appendixA}). 
The deviation observed in the Group B may be attributed to the small size of the central region and the large statistical error. Because the small size of central region contains a small number of stars, even a slight fluctuation in the number of stars from background can lead to a significant amplification in the number density distribution. This amplification may cause the discrepancy between the observed density distribution and the theoretical model. 
In addition, our model assumes spherical symmetry and dynamical equilibrium for clusters. However,  clusters may deviate from spherical symmetry \citep{2008Gieles, 2008Gieles-b, 2019Meingast-b, 2020Zhang, 2022Pang}, or clusters may have not sufficient time to reach dynamical equilibrium \citep{2019Krumholz}. We speculate that it is a secondary factor contributing to the deviation between the observation and our model.
For OCs in Group B, $W_0$ could be misestimated due to the model's inability to accurately reproduce the central regions of the clusters and we don't perform statistical analysis for them.

The table \ref{tab:table1} shows a portion of OCs in Group A. The first column shows the names of OCs. Columns 2 - 7 list the basic physical information of OCs, which are derived from CG20. Columns 8 - 10 list the structural parameters derived from the 50th percentile of the MCMC sampling results, and the error ranges defined by the 16th to 84th percentiles from the MCMC sampling results.

\begingroup
\renewcommand{\arraystretch}{1.5} 

\begin{table*}
\caption{Final parameters summary of OCs in Group A }
  \hspace*{-0.5cm}
     \begin{tabular}{ccccccccccc} 
\hline

Cluster & $RA\_ICRS$ & $DE\_ICRS$ & pmRA & pmDE & plx & log Age & $W_0$ & g & $R_h$ \\

-& (deg) & (deg) & (mas/yr) & (mas/yr) & (mas) & (dex) &- & -& (pc)  \\
\hline
ASCC\ 105 & 295.548 & 27.366 & 1.464 & -1.635 & 1.783 & 7.87 & $4.844^{+1.414}_{-2.613}$ & $1.424^{+0.982}_{-0.983}$  & $6.209^{+1.189}_{-1.089}$ \\
ASCC\ 107 & 297.164 & 21.987 & -0.155 & -5.156 & 1.109 & 7.23 & $3.484^{+2.013}_{-2.088}$ & $2.065^{+0.607}_{-0.965}$ & $2.624^{+0.694}_{-0.527}$ \\
ASCC\ 108 & 298.306 & 39.349 & -0.519 & -1.690 & 0.838 & 8.03 & $5.703^{+0.795}_{-1.402}$ & $0.486^{+0.956}_{-0.377}$ & $13.360^{+1.681}_{-2.123}$ \\
ASCC\ 11 & 53.056 & 44.856 & 0.926 & -3.030 & 1.141 & 8.39 & $4.161^{+1.444}_{-1.765}$ & $1.780^{+0.582}_{-0.706}$ & $5.470^{+0.681}_{-0.516}$ \\
ASCC\ 111 & 302.891 & 37.515 & -1.150 & -1.524 & 1.166 & 8.44 & $6.076^{+0.738}_{-1.059}$ & $0.427^{+0.861}_{-0.306}$ & $9.318^{+1.127}_{-1.555}$ \\
ASCC\ 12 & 72.400 & 41.744 & -0.634 & -2.794 & 0.941 & 7.97 & $5.410^{+1.281}_{-2.192}$ & $1.544^{+0.899}_{-0.991}$ & $4.758^{+1.213}_{-0.946}$\\
ASCC\ 127 & 347.205 & 64.974 & 7.474 & -1.745 & 2.633 & 7.26 & $5.189^{+0.870}_{-2.395}$ & $1.048^{+1.167}_{-0.728}$ & $4.899^{+0.706}_{-0.624}$\\
ASCC\ 16 & 81.198 & 1.655 & 1.355 & -0.015 & 2.838 & 7.13 & $2.499^{+1.454}_{-1.333}$ & $1.419^{+0.460}_{-0.560}$ & $3.216^{+0.251}_{-0.217}$\\
ASCC\ 19 & 81.982 & -1.987 & 1.152 & -1.234 & 2.768 & 7.02 & $5.026^{+1.744}_{-2.725}$ & $1.166^{+1.263}_{-0.871}$ & $3.392^{+0.919}_{-0.789}$\\
ASCC\ 23 & 95.047 & 46.710 & 0.646 & -3.990 & 3.352 & 8.37 & $5.929^{+0.865}_{-1.449}$ & $0.968^{+0.833}_{-0.608}$ & $4.698^{+0.774}_{-0.625}$ \\

 $...$  &  $...$ &  $...$ &  $...$ &  $...$ &  $...$ &  $...$ &  $...$ &  $...$ &  $...$ \\

\hline
					
	\end{tabular}
\vspace{2mm}
\parbox{\linewidth}{\footnotesize
\textbf{Note.} \\
$^{\rm a}$The table only displays a part of OCs' parameters. \\
$^{\rm b}$Columns 2-7 list the basic physical information of OCs, which are from \cite{2020Cantat-Gaudin}. \\
$^{\rm c}$Columns 8-10 list the structural parameters, which are from our model fitting. The optimal parameters are given by the 50th percentile of the MCMC sampling distribution, and the error range defined by the 16th to 84th percentiles.
}

  \label{tab:table1}
\end{table*}

\subsection{Statistical Properties of Intrinsic Parameters}\label{sec:4.1}

We use the structural parameters from the Limepy model and the age from \cite{2020Cantat-Gaudin} to investigate their mutual correlations. The left panel of Figure \ref{fig:2} illustrates the relationship between the cluster structural parameters $W_0$ and $g$, where the colorful dots represent OCs in Group A and are color-coded with the $R_h$. Clusters in Group B are consistently represented by grey triple dots in all  figures and excluded from statistical analysis due to their less reliable results.
The relationship shows that the central concentration $W_0$  tends to be anti-correlated with truncation parameter $g$, indicating that OCs with denser centers exhibit sharper truncations in the outskirts. Similar structural study of GCs (derived from \cite{2019de-Boer}, marked as green dots with error bars) shows a similar negative correlation between $W_0$ and $g$.  GCs occupy a different region than OCs on the $W_0$ vs $g$ diagram, which are generally located in the upper right region. However, there is no clear boundary but a slight overlap between OCs and GCs. We display the distributions of the parameter $W_0$ in the right panel of Figure \ref{fig:2}:  the $W_0$ values of OCs (purple line) concentrate between 2 and 6 while the $W_0$ values of GCs typically gather between 4 and 8 (green line).

We also notice that some clusters located at lower left corner  (indicated within the dashed-line box) in the left panel of Figure \ref{fig:2}, which have both low $W_0$ and $g$ value. Their $R_h$ are significantly larger than the average value of all clusters. We  also highlight them with the red open-circle dots in the Figure \ref{fig:3}, where they occupy the position of large $R_h$ in the diagram.

\begin{figure*}
    \centering
   \includegraphics[width=1.05\linewidth]{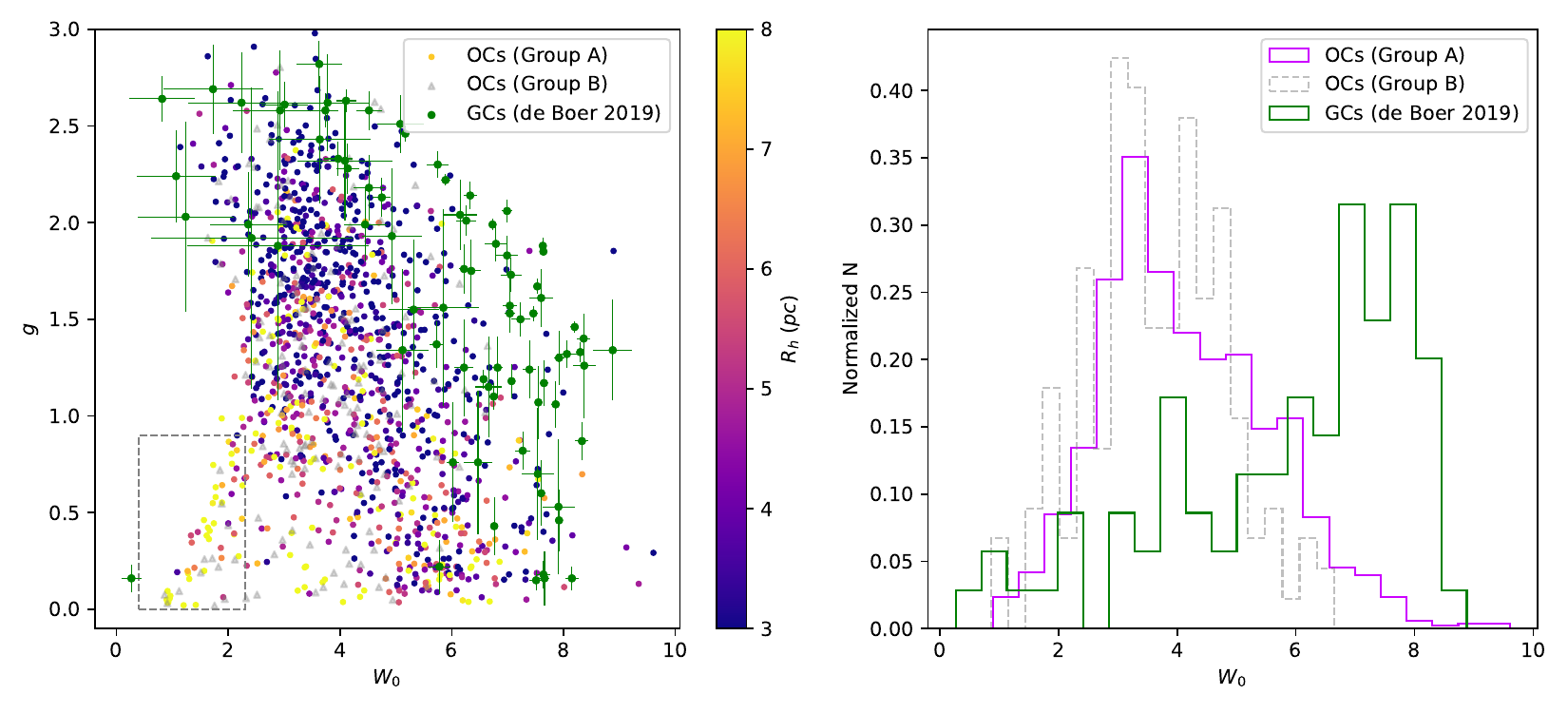}
    
    \caption{(Left panel) The relationship between structural parameters $W_0$ and $g$. The color dots represent OCs in Group A, colored with the half-mass radius $R_h$. The grey triple  dots show the result for OCs in Group B. The green dots stand for GCs, which are from  \protect \cite{2019de-Boer}. The OCs in dashed-line box have low values of $W_0$ and $g$, while their $R_h$ are relatively large. (Right panel) The  histograms represent the distributions of  $W_0$, comparing OCs and GCs  \protect \citep{2019de-Boer}. The purple line represents OCs in Group A, while the green line represents GCs.}
    \label{fig:2}
\end{figure*}

\begin{figure*}
   \hspace*{-0.5cm}
 
    \includegraphics[width=1.05\linewidth]{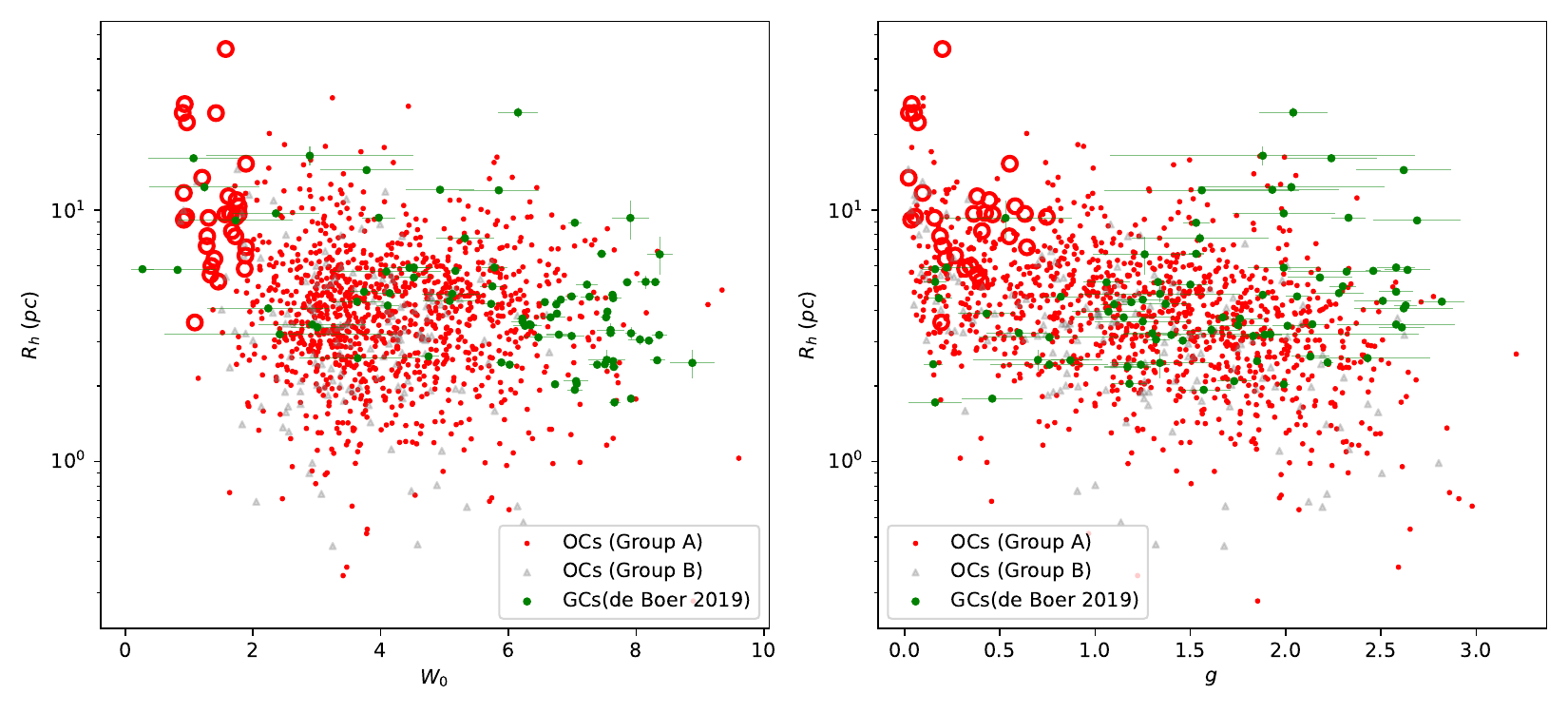}
    
    \caption{((Left panel) $W_0\ vs\ R_h$ between OCs (OCs in Group A for red dots and  OCs in Group B for grey triple  dots) and GCs (green dots). (Right panel) $g\ vs\ R_h$ between OCs (OCs in Group A for red dots and OCs in Group B for grey triple  dots) and GCs (green dots). The red open-circle dots highlight the subset of OCs located within the dashed box in the left panel of Figure \ref{fig:2}, which have low values of $W_0$ and $g$.
     }
    \label{fig:3}
\end{figure*}

The Figure \ref{fig:3} shows the structural parameters $R_h$ with $W_0$ and $g$, where the red dots (solid dots and open-circle dots) represent OCs in Group A and the green dots with error bars represent GCs \citep{2019de-Boer}. The red open-circle dots highlight the subset of OCs located within the dashed box in the left panel of Figure \ref{fig:2}. We find that OCs and GCs exhibit similar structural relationship in $R_h$ vs $W_0$ diagram — 
OCs and GCs with $W_0 > 6$ exhibit that their $R_h$ predominantly concentrate below 10 pc. Figure \ref{fig:3} shows the structural parameters $R_h$ have a great overlap between OCs and GCs, which is consistent with the results of \cite{2010Portegies}.

The structures of clusters are related to factors such as their origin environment and evolution, suggesting that the structural parameters may be related to the age \citep{2005Bonatto,2008Gieles, 2021Krumholz, 2024Alvarez-Baena}. 
The left and right panels of Figure~\ref{fig:4} show the relationships between structural parameters $W_0$, $g$ and the age, and the blue dots represent OCs in Group A. 
The green lines in two panels stand the mean values of $W_0$ and $g$ in different age bins and the light green regions indicate the range of standard deviations of $W_0$ and $g$ binned by $log(Age)$.
The statistical analysis shows that there is a weak correlation between the structural parameters $W_0$, $g$ and the age of the clusters. This observation result implies that the structural parameters $W_0$ and $g$ are not solely determined by cluster age.

We show the relationship between the half-mass radius $R_h$ and $log(Age)$ in the Figure~\ref{fig:5}, where the lower limit of the $R_h$ shows a evident increase with age. Fitting a linear function to the lower limit of the $R_h$ and age derives a correlation of $(R_h)_{low-limit} = 0.567 \times \mathrm{log (Age)} - 3.570$ as shown as a green line in the Figure~\ref{fig:5}.

\begin{figure*}
    \centering
    \includegraphics[width=1\linewidth]{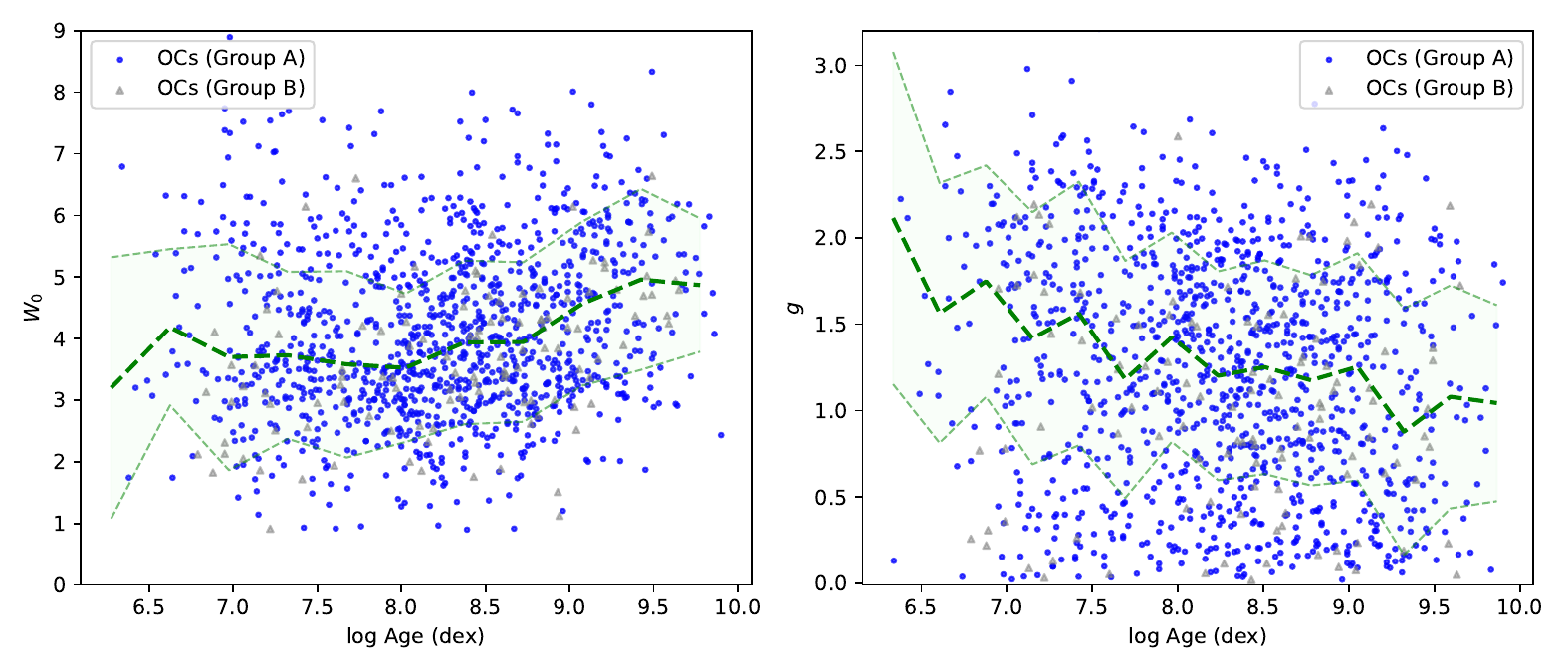}
    \caption{The left and right panels of the figure show the relationships between age and the structural parameters $W_0,\ g$, respectively. Blue points represent OCs in Group A, while grey triple dots indicate  OCs in Group B, which are not included in the statistical analysis. The green lines stand the mean values of $W_0,\ g$ in different $log(Age)$ bins and the light green regions indicate the range of standard deviations of $W_0,\ g$ binned by $log (Age)$. }
    \label{fig:4}
\end{figure*}

\begin{figure}
    \centering
    \includegraphics[width=1\linewidth]{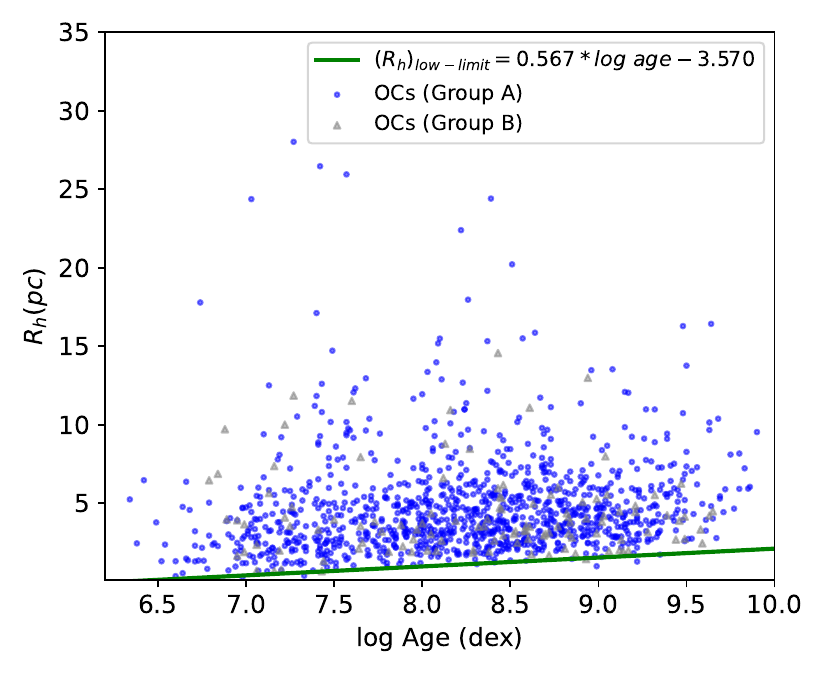}
    \caption{The figure show the relationships between age and the structural parameters $R_h$ in our sample. Blue points represent OCs in Group A, while grey triple dots indicate  OCs in Group B, which are not included in the statistical analysis. The green line is the lower limit of the half-mass radius within different age bins.}
    \label{fig:5}
\end{figure}

\subsection{Spatial Distribution}\label{sec:4.2}

\begin{figure*}
    \centering
    \includegraphics[width=0.8\linewidth]{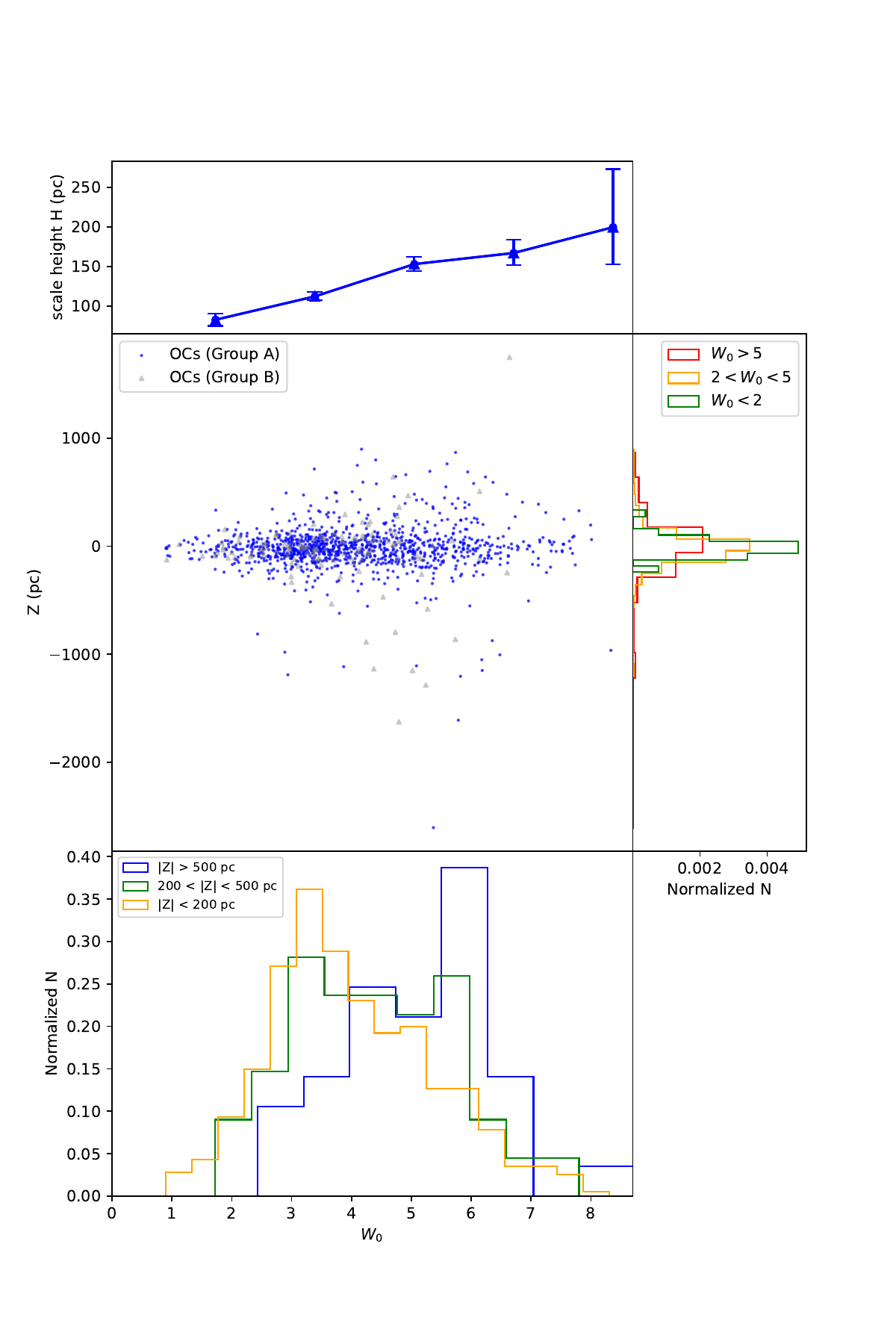}
    \caption{The distribution of $W_0$ in the direction perpendicular to the Galactic plane. Blue points represent OCs in Group A, while grey triple dots indicate  OCs in Group B, which are not included in the statistical analysis. The Z distributions of OCs with $W_0 > 5$ (red line),   $2< W_0 < 5$ (orange line) and $W_0 < 2$ (green line) are shown in the right histogram. The $W_0$ distributions of OCs with $|Z| > 500 \ pc$ (blue line),  $200 < |Z| < 500 \ pc$ (green line) and $|Z| < 200 \ pc$ (orange line) are shown in the bottom histogram. The scale heights for different $W_0$ bins are shown in the upper panel of figure. }
    \label{fig:6}
\end{figure*}

The evolution of the OC is influenced not only by its own internal processes, but also by the interactions with other objects or structures in the Milky Way, such as passages through the Galactic  disk or  encounters with GMCs \citep{2006Piskunov, 2008Binney, 2019Krumholz, 2023Cheng,2024Alvarez-Baena}. OCs at different locations are influenced by different tidal fields strength of the Milky Way  and different interactions with GMCs \citep[e.g.][]{2013Morales}. 
As a result, we would expect that the structural parameters of star clusters are related to their locations within the Milky Way.

Figure \ref{fig:6} shows the distribution of $W_0$ along the Galactic z-axis, defined as the  position perpendicular to the Galactic plane  (labeled as the Height $Z$).
The statistical analysis shows that OCs located far from the disk tend to have larger $W_0$ values, indicating the denser central concentrations. To quantify the vertical distribution of clusters away from the Galactic plane, we use the Scale Height H, which is defined as the parameter in the exponential function $ \exp\left(-\frac{\left|Z\right|}{H}\right) $. We use MCMC method to obtain the scale height H for a set of heights Z. The increased trend is clearly shown in the scale heights calculated within different $W_0$ bins, as OCs with $W_0$ around 1 have a scale height of about $82.11_{-7.21}^{+8.43}$ pc and OCs with $W_0$ around 8 have a scale height of about $196.56_{-44.70}^{+63.50}$ pc in the top panel of Fig \ref{fig:6}. In addition, we divide the OCs into groups with $W_0 > 5$ (red line), $2<W_0 < 5$ (orange line) and $W_0 < 2$ (green line) for the histogram in the right panel of Fig \ref{fig:6}, from which we find that the scale heights are $180.14_{-10.38}^{+10.75} \ pc$, $111.82_{-3.76}^{+4.12} \ pc$ and $70.36_{-9.76}^{+11.82}\ pc$, respectively. 
Another difference presents when we divide our sample into three subsamples based on their distance from the Galactic plane in the bottom histogram of Fig \ref{fig:6}. The $|Z| < 200 $ pc subsample exhibits a  peak in $W_0$ at approximately 3. In contrast, the $200 <|Z| < 500$ pc subsample shows a relatively flat distribution across $W_0$ at 3 – 6, while the $|Z| > 500$ pc subsample shifts toward larger $W_0$ with a peak at near 6.


\begin{figure*}
    \centering
 \includegraphics[width=1\linewidth]{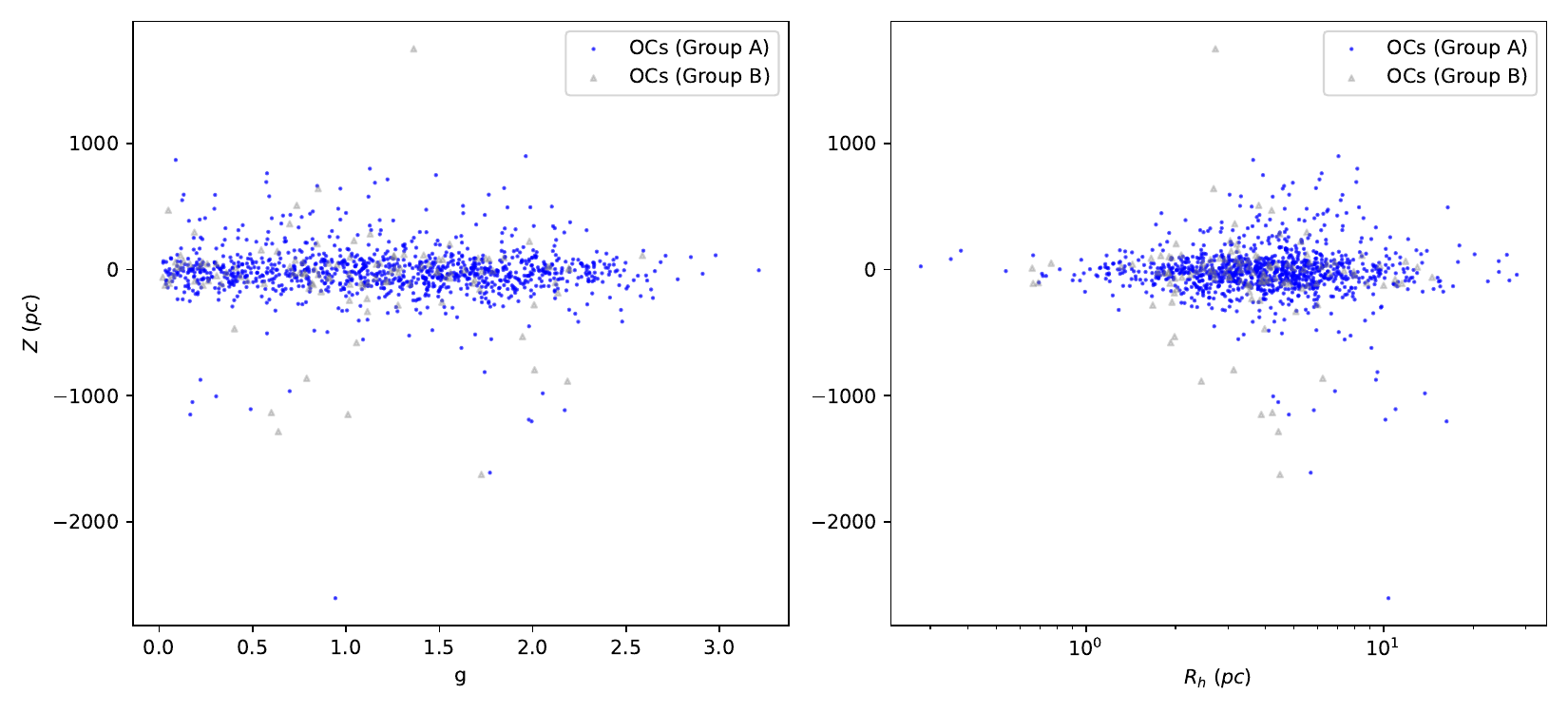}
    \caption{The relationship between the structural parameters $g$ (Left-panel) and $R_h$ (Right-panel) with height Z. Blue points represent OCs in Group A, while gray triple points indicate OCs in Group B.}
    \label{fig7}
\end{figure*}

We also study the relationship of the height $Z$ with $g$ and $R_h$ in Figure \ref{fig7}, which  shows there is no simple correlation between them. It may imply $g$ and $R_h$ are influenced by more factors, and do not show a clear simple relationship with height Z. 

In addition to the vertical distribution, we also examine how structural parameters vary with galactocentric distance $R_{gc}$. As shown in the Figure \ref{fig:8}, $W_0$ exhibits a weak trend of increasing with $R_{gc}$, suggesting that clusters at larger distances from the Galactic center tend to be slightly more concentrated. In contrast, no significant trends are observed for the truncation parameter $g$ or the half-mass radius $R_{h}$ as functions of $R_{gc}$.

\begin{figure*}
    \centering
 \includegraphics[width=1\linewidth]{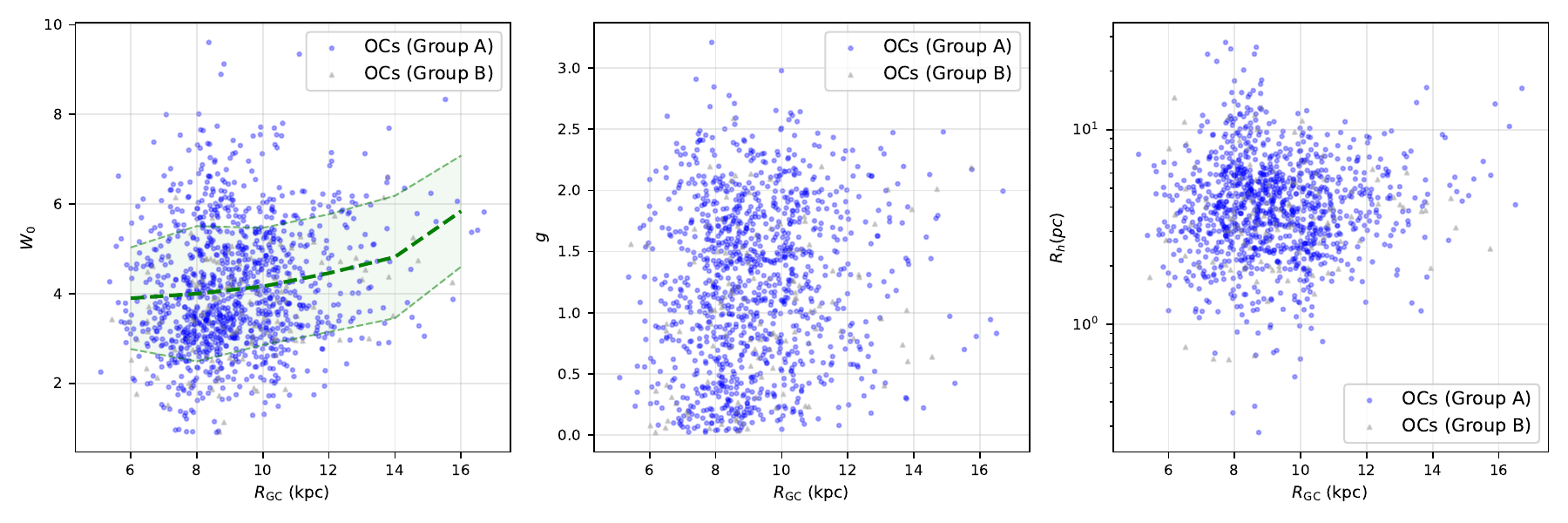}
    \caption{The relationship between the structural parameters $W_0$ (Left- panel), $g$ (Middle-panel) and $R_h$ (Right-panel) with $R_{gc}$. Blue points represent OCs in Group A, while gray triple points indicate OCs in Group B.}
    \label{fig:8}
\end{figure*}

\section{Conclusion And Discussion} \label{sec:5}

In this study,  the  Limepy model combined with uniformly distributed field stars model is used to fit the surface distribution of a large sample of 1,481 OCs. The structural parameters are obtained by maximizing the likelihood function of our model using MCMC methods. We obtain the structural parameters of 1,160 OCs, and the corresponding models show good agreement with the observational data.

Our analysis reveals a negative correlation between the structural parameter $W_0$ and $g$, which shows that OCs with denser central  concentrations tend to have the sharper truncations at the outskirts of clusters. This observed trend suggests a possible evolutionary link between the internal structure and the dynamics of star clusters. To better understand such structural correction, numerical simulations offer valuable insights into how cluster structure changes over time under different initial conditions.
For instance, \cite{2016Zocchi} conducted an $N$-body simulation on a cluster comprising 65,536 stars. Their results indicated that $W_0$ increases  from an initial value of 4 to a peak of about 11, and slightly decreases in the later phases. The structural parameter $g$ decreases  continuously over the process of evolution and the $g$ value decreases from 2.5 to 0.5 for the simulated cluster. The underlying physical mechanisms can be attributed to two main factors. Mass segregation in the cluster causes high mass stars (as well as multiple stellar systems) to sink towards the center and become more concentrated in the core as the cluster evolves \citep{2005Bonatto, 2022Llorente}. The continuous stripping of the cluster's outer stars by the tidal field leads to a more sharply truncated outer profile \citep{2016Zocchi}. These findings imply that the structural evolution of star clusters may be shaped by a combination of internal dynamical processes and external environmental influences. The statistical anti-correlation between $W_0$ and $g$ likely reflects that most star clusters evolve along a similar structural evolutionary path. We also examined the overall correlations between $W_0$ and age, as well as between $g$ and age. The results show weak correlations in both cases. This suggests that individual structural parameters may weakly correlate with age, likely due to the diverse dynamical properties of clusters and the varying environmental influences they experience.  

Our result shows that the lower limit of cluster $R_h$ increases  with age. 
The increasing trend may result from  the cluster expansion during the evolution. The cluster's mass loss (e.g., two-body relaxation, tidal stripping)  and the energy gain (e.g., the encounters with GMCs) can 
 reduce the binding energy of the cluster and cause the cluster expansion  \citep{2006Gieles,2019Webb,2019Krumholz}. In addition, the smaller clusters are  more vulnerable to the effects of the  evolution progresses and make it  more difficult for them to survive longer time, which may cause the scarcity of low $R_h$ among the old OCs \citep{2006Gieles,2010Gieles}.  

The structural parameter $W_0$ tends to be higher for the OC located at a greater distance from the Galactic disk. 
The OCs located farther from the Galactic disk  experience a strong gradient variation of tidal field  during the disk crossings, making them more likely to be disrupted by the large vertical tidal shocks \citep{2017Martinez, 2019Krumholz}.  It implies that OCs  farther from the Galactic disk require more stable structure to survive during experiencing stronger tidal shocks, indicating they tend to have denser central concentrations.

There is a overlap of structural parameters between OCs and GCs without a clear boundary or gap.
The overlap between OCs and GCs has been discovered in previous studies, such as   the half-mass radii \citep{2010Portegies,2019Krumholz}. 
Besides, OCs and GCs exhibit the similar negative correlation between $W_0$ and $g$.  
The similarity of structural correlation  may imply that OCs and GCs   experience similar physical dynamical processes, despite the differences in OCs' and GCs' typical masses, stellar populations and locations in the Milky Way.

In summary, this work presents a structural analysis of a large sample of OCs, revealing key correlations between their structure and basic physical parameters. These findings may offer insights into the evolutionary processes of star clusters' structure influenced by internal dynamics and external conditions.

\section*{Acknowledgements}

This work is supported by the Strategic Priority Research Program of the Chinese Academy of Sciences (grant NO. XDB0550300). ZW acknowledge the support of the China Postdoctoral Science Foundation (2022M723059); the China Postdoctoral Science Foundation (2023T160615); the Youth Innovation Fund (WK2030000080). X.Z.L acknowledge the support of the Anhui Provincial Natural Science Foundation (2308085QA35).
This work is supported by National Key Research and Development Program of China (2023YFA1608100). The authors appreciate the support of the National Natural Science Foundation of China (NSFC, Grant Nos. 12173037, 12233008), the CAS Project for Young Scientists in Basic Research (No. YSBR-092), the Fundamental Research Funds for the Central Universities (WK3440000006), and the Cyrus Chun Ying Tang Foundations.
This work has made use of data from the European Space Agency (ESA) mission Gaia (https://www.cosmos.esa.int/gaia), processed by the Gaia Data Processing and Analysis Consortium (DPAC, https://www.cosmos.esa.int/web/gaia/dpac/consortium). Funding for the DPAC has been provided by national institutions, in particular the institutions participating in the Gaia Multilateral Agreement.

\section*{Data availability}

The research presented in this article predominantly relies on data
that is publicly available and accessible online through the Gaia DR3 \citep{GaiaCollaboration2023}. The derived structural parameters of the Galactic open clusters analyzed in this work may be shared on reasonable
request to the corresponding author.



\bibliographystyle{mnras}
\bibliography{my} 




\appendix

\section{Examples of OCs with poor fits}\label{appendixA}
OCs in regions of inhomogeneous field star density are excluded because they violate the assumption of uniform background in our model and the obtained structural parameters have no physical meaning. As an example, we show the result of Loden 1194 in the left panel of Figure~\ref{fig:9}. For the remaining OCs showing significant deviations from the observations, the discrepancies between observations and the models are primarily in the centers of the
clusters. We use a simple, artificial criterion: $|Observation - Model|\ /\ Observation > 0.25$ for the center bin to select them and name them Group B. We take ASCC 123  as examples in the right panel of Figure~\ref{fig:9}. 

\begin{figure*}
   
    \centering
    \includegraphics[width=1\linewidth]{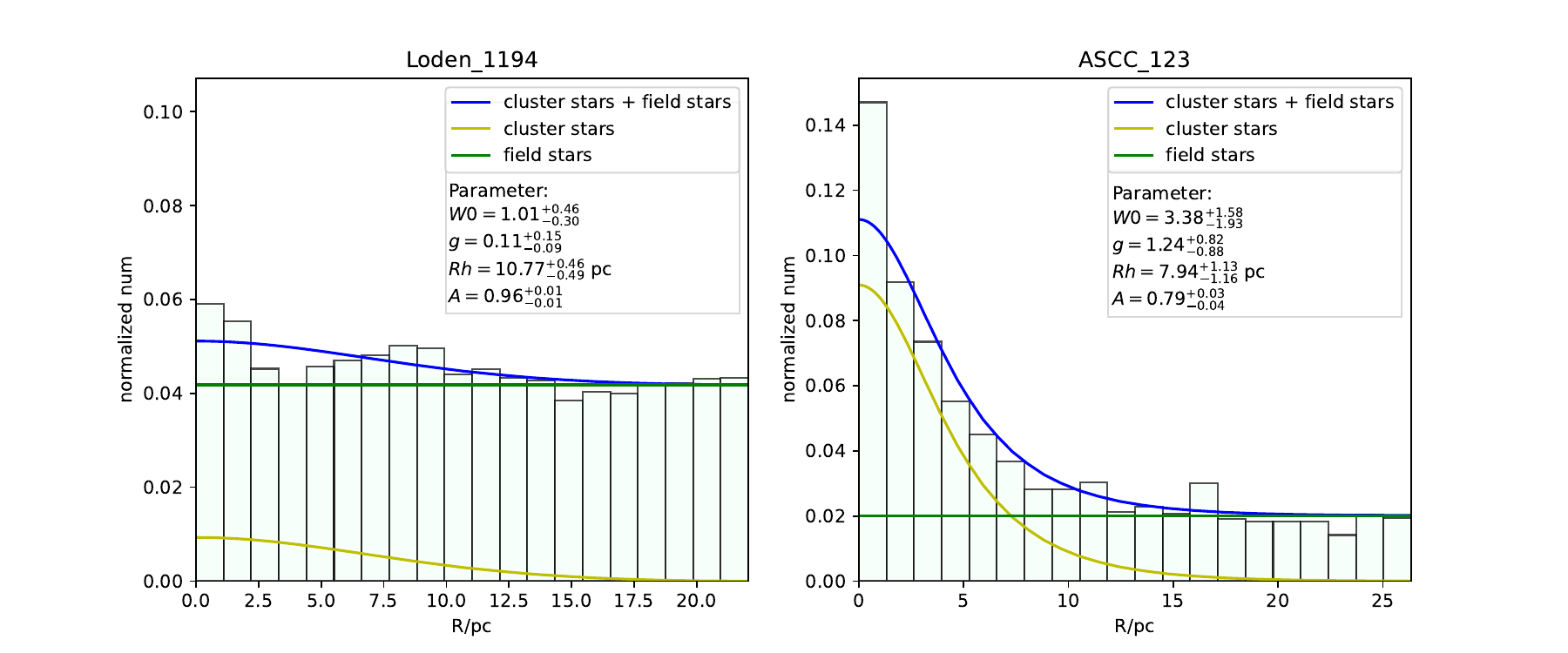}
    \caption{The compared results between the model obtained from MCMC and the original data: Loden 1194 (Left panel) and ASCC 123 (Right panel). The histogram is original data translating into the number density distribution. The blue lines are the model with the best parameters from MCMC. The yellow line and green line represent cluster members and field stars, respectively.}
    \label{fig:9}
\end{figure*}


\bsp	
\label{lastpage}
\end{document}